\documentstyle{article}

\def\be{\begin{equation}}
\def\ee{\end{equation}}
\def\eel#1 {\label{#1}\end{equation}}
\def\rz#1 {(\ref{#1}) } \def\ry#1 {(\ref{#1})}
 \def\ba{\begin{eqnarray}}
 \def\ea{\end{eqnarray}}
 \def\la{\label}
\def\nn{\nonumber \\}

\let\a=\alpha \let\b=\beta \let\g=\gamma \let\d=\delta
\let\e=\varepsilon \let\ep=\epsilon  

   \let\m=\mu
    \let\s=\sigma
 \let\o=\omega  
   
\let\O=\Omega \let\S=\Sigma  
 \let\G=\Gamma  
\def\CX{{\cal X}}\def\CC{{\cal C}}\def\CS{{\cal S}}
 \def\CP{{\cal P}}\def\CR{{\cal R}}  
\def\0{\over } \def\1{\vec } \def\2{{1\over 2}} \def\4{{1\over 4}}
\def\5{\bar } \def\6{\partial }
\def\7{\mbox{$\2$}}
\def\8{\mbox{${1 \0 4}$}}

\def\({\left(} \def\){\right)} \def\<{\langle } \def\>{\rangle }
\def\[{\left[} \def\]{\right]}  
\let\lra=\leftrightarrow 
\let\Ra=\Rightarrow \let\ra=\rightarrow

 \newcommand{\dR}{\mbox{{\sl I \hspace{-0.8em} R}}}
 
\newcommand{\done}{\mbox{{\sl {\bf {\sl 1}} \hspace{-1em} I}}}

\begin{document}

\title{{\bf A Brief Introduction to Poisson $\sigma$-Models}}

\author{{\bf PETER SCHALLER}\thanks{e-mail: schaller@tph.tuwien.ac.at
\newline
\hspace*{8 pt}
$^\diamondsuit$e-mail: tstrobl@pluto.physik.rwth-aachen.de}
\\
Institut f\"{u}r Theoretische Physik, TU Wien
\\
Wiedner Hauptstr. 8-10, A-1040 Vienna, Austria
\\[4mm]
{\bf THOMAS STROBL}$^\diamondsuit$
\\
Institut f\"{u}r Theoretische Physik, RWTH Aachen
\\
Sommerfeldstr. 26-28, D-52056 Aachen, Germany
\\ \\ }

\vspace{0.6cm}
\date{based on two talks delivered in Schladming, March 1995}
\vfill
\maketitle 
\vspace*{-8.3cm}
\hspace*{7cm}
{\large \tt  TUW - 95 - 15 \ \ PITHA - 95/15} \\
\hspace*{9.3cm}
{\large \tt hep-th/9507020}
\vspace*{7.5cm}

The close interplay between geometry, topology and algebra turned out
to be a most crucial point in the analysis of low dimensional field
theories. This is particularly true for the large class of topological
and almost topological two-dimensional field theories which can be treated
comprehensively in the framework of Poisson-$\sigma$ models. Examples
are provided by pure Yang-Mills and gravity theories and, to some
extent, by the G/G gauged WZW-model.

It turned out \cite{Peter,Anton,proce} that the common mathematical
structure behind these theories is the one of
Poisson manifolds. In the present contribution we review  how this
structure leads to the formulation of the Poisson-$\sigma$ models. We
will demonstrate that the general features of Poisson manifolds
provide powerful tools for the analysis of these models. And we will
apply the results to the examples mentioned above.

For pedagogical reasons we will, in the presentation of the
examples, restrict ourselves to $SU(2)$ as gauge groups of
YM and G/G and, in the case of gravity theories,
to ${\cal R}^2$-gravity ($\CR$ being the Ricci scalar).
This has the advantage that the considered target
spaces will turn out to be three-dimensional then, opening possiblities
for a simple visualization and  explicitness of formulas.
The method is, however, by no means restricted to these cases; for YM
and G/G,
e.g., even arbitrary non-compact gauge groups  are comprised by the
treatment.

Let us start by reviewing some features of Poisson manifolds \cite{Mad},
which
finally will become the target space of our 2D field theories. (To avoid
any confusion, let us note
right away that the notion of a Poisson manifold is somewhat
 more general than the one of a symplectic manifold).

Denote by $N$ a manifold equipped with a Poisson bracket relation
$\{.,.\}$ between functions on $N$. By definition the Poisson bracket
is bilinear and antisymmetric. It is, moreover, subject to the
Leibnitz rule $\{f,gh\}=\{f,g\}h
+g\{f,h\}$ and the Jacoby identity $\{f,\{g,h\}\}+cycl.=0$.

In local
coordinates $X^i$ on $N$ a Poisson bracket may be expressed
in terms of a two-tensor $\CP^{ij}(X):=\{X^i,X^j\}$.
With the Leibnitz rule this relation determines the Poisson
structure  uniquely:
   \be
\{f,g\}=\CP^{ij}f_{,i}g_{,j} \, .
   \eel Tens
where $f_{,i} \equiv \6 f/ \6 X^i$ and summation convention is
understood. From \rz Tens it is obvious that $\CP^{ij}$
transforms covariantly under coordinate transformations. In terms of
$\CP^{ij}$ the Jacoby identity reads
\be
  \CP^{ij}{,_k}\CP^{kl}+cycl.(i,j,l)=0 \, .
\eel Jaco

There is another consequence of the Leibnitz rule: The Poisson bracket
of the constant function with anything else vanishes. In the case of a
general Poisson manifold there may be also other functions which
share this property. They are called Casimir functions.

An example for a $2n$-dimensional Poisson manifold is provided by the
phase space of an $n$-particle system: Denote by $Q^\alpha$ the positions of
the particle and by $P_\alpha$ the momenta. The Poisson structure is then
given by $\{Q^\a,P_\b \}=\delta^\a_\b$ and $\{Q^\a, Q^\b \} = 0 =
\{P_\a,  P_\b \}$.

With \rz Tens it is easy to see that   the constant function is the only
Casimir function in this example, which obviously holds whenever
$\CP^{ij}$ is
nondegenerate. Actually
Poisson manifolds with nondegenerate $\CP$ are symplectic
manifolds,  the symplectic two-form being equal to the inverse
of the Poisson tensor $\CP$.  For them the above example is in some
sense generic: Locally {\em any} symplectic manifold allows
so-called Darboux coordinates such that the Poisson tensor takes
the form of the $2n\times 2n$ matrix:
\be
  \left(
  \begin{array}{rr}
  0&\done\\-\done&0
  \end{array}
  \right) \quad ,
\eel Darb
where $\done$ denotes the $n \times n$ unit matrix.

Another example for a Poisson structure is given by the Poisson
bracket
\be
  \{X^i,X^j\}=\sum_{k=1}^3 \varepsilon^{ijk} X^k
\eel Brac
 on $\dR^3$, where $\varepsilon^{ijk}$ is the completely antisymmetric
three-tensor.
Obviously the Poisson structure \rz Brac is degenerate:
Darboux coordinates cannot exist on a manifold of odd dimension.
Indeed $R^2:= \sum_{i=1}^3 X^iX^i$ is a Casimir function.

In the example \rz Brac the Poisson structure can be
restricted consistently to any two-sphere given by the choice of a
constant value for $R^2$. The restricted Poisson structure is
nondegenerate. It is easy to convince oneself that Darboux coordinates
on a two-sphere of radius $R_0$ are provided by one of the coordinates
$X^i$ and the azimutal angle around the
respective axis; e.g.\ with $Z :=  X^3$ and $\Phi := \arctan (X^2/X^1)$
it follows from  \rz Brac that $\{ \Phi , Z \}=1$.
The physical phase spaces corresponding to these two-spheres are the
classical analogues of spin systems. (We will come back to this point
below.)

The picture outlined here generalizes in a straightforward way:
A Poisson manifold with degenerate Poisson structure  foliates into a
family of lower
dimensional manifolds (symplectic leaves) each of which is characterized
by assigning constant values to the Casimir functions and is equipped with a
nondegenerate Poisson structure. To be more precise: Some of the symplectic
leaves may have lower dimension than the generic ones, such as the
origin $X^i=0$ in the example \ry Brac , and  in some of these
cases (not so for
\ry Brac ) level surfaces of Casimir functions correspond to several
symplecitc leaves.
Moreover, a symplectic leaf need not be a (regular) submanifold of $N$,
i.e.\  it might  wind around densely in some higher dimensional
subregion of $N$ (as a higher-dimensional generalization of what
one knows from paths on a torus in classical mechanics, cf.,
e.g., \cite{Thi}). In that case constant values of Casimir functions
characterize parts of such  leaves  only.  So what we are dealing with
is a stratifacation of $N$
rather than a foliation. Nevertheless we will stick
 to the more
common nomenclature introduced before.

As an example for a nonlinear Poisson structure let us take the
quadratic modification
\be \{X,Y\}= - Z^2 + {1 \0 4} \; , \quad \{ Y, Z \} = X \; , \quad
\{Z, X \} = Y  \eel PR2
of \rz Brac on $\dR^3$ with coordinates $(X,Y,Z)$, which will become
relevant
in the analysis of (Euclidean) $\CR^2$-gravity (with cosmological constant).
It is straightforward to check that
\be C_{\CR^2}:= X^2 + Y^2 - \mbox{${2 \0 3}$} Z^3 + \7 Z  \, , \eel CR2
 is a Casimir function of the bracket \ry PR2 .
$\Phi := \arctan (Y/X)$ and $Z$, on the other hand,  are still canonical
conjugates (i.e.\ $\{\Phi, Z\}=1$). They may be used as
local Darboux coordinates on any of the
symplectic  leaves, which (generically) coincide with the
level surfaces of \ry CR2 .

It is a nice excercise to visualize the 'foliation' of $N=\dR^3$ into
symplectic leaves. This is done most easily by rotating the square
 root of the positive parts of
\be h(Z) :=  \mbox{${2 \0 3}$} Z^3 - \7 Z+ C_{\CR^2} \eel h
around the $Z$-axis (since $h(Z) = X^2+Y^2$). The resulting
picture is the following: For values of $C_{\CR^2}$ smaller than $-1/6$
the symplectic leaves are diffeomorphic to the $(X,Y)$-plane. At $C_{\CR^2}
=-1/6$ there is an additional pointlike leaf at $(X,Y,Z)=(0,0,-\7)$.
Indeed, this is one of the two points where the right-hand side of \rz PR2
(and thus also the Poisson tensor $\CP$) vanishes. Within the range
$(-1/6,1/6)$ of the Casimir, the pointlike leaf turns into an
ellipsoid, again accompanied by a 'plane' (situated at larger values
of $Z$).
For $C_{\CR^2}=1/6$ the ellipsoid and the plane touch
at $(X,Y,Z)=(0,0,\7)$. This value of $C_{\CR^2}$ corresponds to
three symplectic leaves: An ellipsoid with one pole taken out ($\sim$
plane),
this pole $(X,Y,Z)=(0,0,\7)$, and a 'plane without origin'.
For larger values of the Casimir, finally, one has only  planar leaves
again.

Let us come back to the Poisson structure \rz Brac
and see its  role in
the two dimensional Yang Mills theory. The field content of the latter
is given by an $su(2)$ algebra-valued one-form $A=A_iT^i$ on a
two-dimensional
worldsheet $M$, where  the generators $T^i$ of $su(2)$  may be
represented by  imaginary multiples
of the Pauli matrices. Then the action reads:
\be
   L_{SU(2)} = {1\over 2\alpha} \int_M tr \, F \wedge \ast F\, ,  \quad
   F=\left(dA_i+\mbox{$\2$}{\varepsilon^{jk}}_iA_j\wedge A_k\right)T^i
\, .
\eel Yang
Here $\alpha$ is a coupling constant and the Hodge dual $*$
is to be taken with respect to some volume-form $\mu$ on $M$.

To analyze the model one may find it useful to rewrite the action
in a first order formalism. Denoting the momenta conjugate to the
connection $A$ (or, more precisely, to $A_{1i}$) by $X^i$, one
obtains
\be
   L'_{SU(2)}=\int_{M}X^i(dA_i+
   {1\over 2} {\varepsilon^{jk}}_iA_j\wedge A_k) -
   {\alpha \0 2} \mu X_iX^i\, .
\eel Yang2
The equivalence to \rz Yang is seen most easily  by inserting
 the field equation $X_i={1 \0 \alpha} *F_i$ into   \ry Yang2 .
Equivalently one might perform the Gaussian integral over $X^i$
in \ry Yang2 . Now with \rz Brac and a partial integration
the $SU(2)$-YM-action \rz Yang2 may be seen to be of the form
\ba   L&=&L_{top} + L_{\CC}  \nn
L_{top}&=&\int_{M}A_i\wedge dX^i+\mbox{${1\over 2}$}
\CP^{ij}(X)A_i\wedge A_j  \la{acti2} \\
L_{\CC}&=& \int \mu \; \CC(X) \, ,
\nonumber \ea
where $\CC(X)$ denotes the Casimir function $-{\alpha \0 2} R^2$ of the
Poisson bracket \ry Brac .

This form provides a natural generalization of the Yang
Mills theory: Let $M$ be the two-dimensional worldsheet of a
field theory.  As dynamical fields we first take coordinates $X^i$ on
some arbitrary Poisson manifold $N$. Additionally we take a field $A$
which is one-form valued on $M$ as well as on $N$, i.e., with
coordinates $x^\mu$ on $M$, $A$ can be written as $A=A_i \wedge
dX^i= A_{\m i}dx^\m \wedge dX^i$. Then $L_{top}$
provides (the topological part of) the Poisson $\s$-model action.
With the additional input of a volume-form
$\m$ on $M$ and the choice of a Casimir function $\CC$
it may be extended further by the addition of $L_{\CC}$.

$L_{top}$ is defined without any reference to a background metric on $M$.
This establishes that the model defined by the action $L_{top}$ is a
topological field theory of the Schwarz type \cite{Blau}.
In the SU(2)-case this corresponds to the topological
Yang-Mills theory with coupling constant $\alpha \to 0$ (BF-theory).
The addition of a nontrivial $L_{\CC}$ to the action, of course, spoils the
topological nature of the model.

The symmetries of the Poisson-$\sigma$-model \rz acti2 are
a straightforward generalization of the SU(2) gauge symmetry,
where the structure constants are replaced by
derivatives of the Poisson tensor:
\be
 \d_\ep X^i =  \ep_i(x) \CP^{ij} \, , \quad
\d_\ep A_i = d\ep_i + {\CP^{lm}}_{,i} A_l \ep_m \, .
\eel symme
Proving the invariance of the action under these transformations, the
Jacoby identity \rz Jaco is most important. So here it becomes
clear that this identity plays a crucial role in the theory of
Poisson-$\sigma$-models.

{}From the action $L=L_{top} + L_{\CC}$
 the equations of motion follow immediately:
 \be
 \begin{array}{c}
  dX^i +  \CP^{ij} A_j=0 \, , \\[2pt]
  dA_i + \2 \CP^{lm}{}_{,i} A_l \wedge A_m + \CC_{,i}=0 \, .
  \end{array}
\eel equom

In the beginning of this talk we asserted that the special
features of Poisson structures greatly facilitate the analysis of
Poisson-$\sigma$-models. This is best illustrated by the
classification of the local solutions of the field equations:

Remember that any
Poisson manifold foliates into a family of symplectic leaves. The
latter may be parametrized by a complete set of independent
Casimir functions $X^I$. On each of the symplectic leaves we may find
Darboux coordinates $X^\alpha$. Together Casimir functions and Darboux
coordinates give rise to a coordinate system $(X^I, X^\alpha)$ on $N$
with $\CP^{Ii}=0={\CP^{ij}},_{k}$ and $\CP^{\alpha\beta }$ having the
standard form \ry Darb .

In these coordinates any Casimir function ${\CC}$ on $N$
depends on the $X^I$ only.
The equations of motion \rz equom simplify to
\be
 dX^I=0 \, , \quad
 dA_I = -\CC_{,I} \, , \quad
 A_\alpha = (\CP^{-1})_{\beta\alpha} dX^\beta \, ,
\eel eomcd
where, as a matrix,  $(\CP^{-1})_{\beta\alpha} = \CP^{\a\b}$ according to
\ry Darb . So the
$X^I(x)$ have to be constant on $M$, but otherwise arbitrary, whereas
the $X^\alpha(x)$ remain completely undetermined by the field
equations. Any choice of the latter determines $A_\alpha$ uniquely through
the last equation in (\ref{eomcd}). Each of the $A_I$ is determined up
to an exact one-form only.

Still one has not made use of the gauge freedom. As is obvious from
\rz symme any choice of the $X^\alpha$ is gauge equivalent,
and also $A_I \sim A_I
+ dh_I$, where the $h_I$ are arbitrary functions.
Thus locally any solution to the field equations is uniquely
determined by the constant values of the $X^I$.

Let us illustrate the usefulness of such considerations at the example
of 2D $\CR^2$-gravity with Lagrangian
\be L_{\CR^2}= \8 \int_M d^2x \sqrt{\det \, |g|} \; \left(\8 \CR^2 +1 \right)
\;.  \eel R2acti
Here $g$ is the metric on $M$,
which, for simplicity, we take to be of Euclidean signature, and $\CR$ is
the Ricci scalar of the corresponding Levi-Civita connection. First we
have to clarify how the action \rz R2acti fits into our framework.
The decisive  observation is that
\be L'_{\CR^2}= \int X\,(de^1 - \o \wedge e^2) +
Y\,(de^2 + \o \wedge e^1) + Z d\o  + (\8 - Z^2) \; e^1 \wedge e^2
\eel R2acti2
constitutes an equivalent formulation of
\rz R2acti in an Einstein-Cartan formalism with zweibein $e^1,e^2$
(so $g=e^1 e^1 +  e^2 e^2$) and connection one-form $\o$.
Indeed $X$ and $Y$ are seen to be Lagrange multipliers for torsion zero and
the elimination of $Z$, analogous to the one of the  $X^i$ in the transition
from \rz Yang2 to \ry Yang , reproduces the $\CR^2$-term due to the relation
$\CR=2 \ast d \o$. Now it is straightforward to verify that
upon the identification $X^i=(X,Y,Z)$ and  $A_i=(e^1,e^2, \o)$ \rz R2acti2
takes the form \rz acti2 with $\CC=0$ and the Poisson bracket \ry PR2 .

According to our study of that bracket Casimir-Darboux coordinates
are provided here by
$(C_{\CR^2}, \Phi, Z)=: \widetilde X^i\equiv (X^I,X^\a)$
(cf.\ Eq.\ \ry CR2 ). Then \rz eomcd yields
$A_{\widetilde 1}=df$, $A_{\widetilde  2}=dZ$, and $A_{\widetilde
  3}=-d\Phi$, where $f$, $Z$, and $\Phi$ are some arbitrary functions
on the worldsheet $M$. Let us choose $f$ and $Z$ to coincide with
coordinates $x^0/2$ and $x^1$ on $M$, respectively. It then is a simple
calculation to verify that with $A_i={\6 \widetilde X^j \0 \6 X^i} \,
A_{\widetilde j}$  one obtaines for  the metric $g=A_1A_1 + A_2A_2$:
\be g= h(x^1) (dx^0)^2 + {(dx^1)^2 \0 h(x^1)} \eel g
where the function $h$ has been defined in \ry h .
A  procedure completely analogous to
the one above reveals that \rz g holds also for Minkowskian
signature, if one merely changes the sign in front of the first
term. So in the Euclidean and in the Minkowskian case of
\rz R2acti
{\em locally} $g$ is of a generalized Schwarzschild form and is
parametrized by one quantity $C_{\CR^2}$, which plays the role of
the Schwarzschild mass.

Additional structures evolve, if global aspects are
taken into account. In \cite{Klo}, e.g.,
 a classification of all  global  solutions to \rz R2acti
(and many other gravity models)  is provided for Minkowskian
signature. This classification contains
continuous and discrete parameters in addition to $C_{\CR^2}$.
But already the global classification of  Euclidean solutions of
\rz R2acti is still an open problem to the best of our knowledge.
It is hoped that the Poisson $\sigma$-formulation is helpful also in
this respect.\footnote{Actually, it turns out that a classification of
all solutions to the Poisson $\s$-model with Poisson
structure \rz PR2 and trivial fibration (cf.\ below)
on arbitrary Riemann surfaces $M=\S_\g$ is not too difficult. However,
since  $\S_\g$ with genus $\g \neq 1$  has a nontrivial tangent
bundle, $X^1$ and $X^2$ are no more mere
functions on $M=\S_\g$, but they are sections of a nontrivial
vector-bundle. So, in order
to describe Euclidean $\CR^2$-gravity correctly by its
powerful reformulation as a Poisson $\s$-model, a nontrivial
fibration of the target space $N$ over the base-manifold $M$ will have to be
taken into account.}

The features of Poisson manifolds are not only helpful in the
classical theory, but also in the quantum theory of
Poisson-$\sigma$-models. To illustrate this, let us analyze the
quantum states in a Hamiltonian formulation.

To this end let us assume $M$ to be of
the form $S^1\times \dR$, parameterized by a 2$\pi$-periodic coordinate
$x^1$ and the evolution parameter $x^0$.
As the action \rz acti2 is already in first order form, the
derivation of the corresponding Hamiltonian system is simple. The zero
components $A_{0i}$ of the $A_i$  play the role of Lagrange
multipliers giving rise to the system of first class constraints
($\6=\6 /\6 x^1$)
\be
 G^i\equiv \6X^i +  \CP^{ij} A_{1j} \approx 0 \, .
\eel const
The fundamental Poisson brackets (not to be confused with the
Poisson brackets on the target space $N$) are given by
\be
  \{ X^i(x^1), A_{1j}(y^1) \} = - \d^i_j \, \d(x^1-y^1)
\eel poibr
and the Hamiltonian reads
\be
  H= \int dx^1 (\CC  - A_{0i}G^i) \, .
\eel hamil

To quantize the system in an $X$-representation we consider quantum
wave functions as complex valued functionals on the space $\Gamma_N$
of parameterized smooth loops in $N$:
\be
 \Gamma_N=\{ \CX : S^1\to N,x\to X^i(x) \} \, .
\eel loops
Following the Dirac procedure, only such quantum states are
admissible which satisfy the quantum constraints ($\hbar := 1$)
\be
 \hat G^i(x) \Psi[\CX ] = \left(
 \6 X^i(x) + i \CP^{ij}(X) {\d \0 \d X^j(x)} \right)
 \Psi[\CX ] =0 \, .
\eel quanc

To get some insight into the problem of solving \ry quanc , let us
specify the constraints for the SU(2) case, using the Casimir-Darboux
coordinates $(R,\Phi ,Z)$. We have
\be
  (\6 R) \, \Psi =0 \; , \qquad  (\6 \Phi +i {\d \0 \d Z(x)})\Psi =0  \; ,
   \qquad  (\6 Z - i {\d \0 \d \Phi (x)})\Psi =0  \; .
\eel YMcon
The first equation \rz YMcon is seen to restrict the support of
$\Psi$ to loops which are contained entirely in some of the
two-spheres with constant $R$. It is easy to see, furthermore,  that
the remaining equations  \rz YMcon  are solved by
\be \Psi = \exp\left(i\oint_\CX \6 \Phi(x) Z(x) dx\right)  \qquad
\mbox{(locally on $\G_{\CS}$) ,} \eel sol
where $\G_{\CS}$ is the space of loops on a chosen two-sphere
$\CS$. As the differential
equations on  $\G_{\CS}$ are of first order, this local solution
is unique up to a multiplicative (generally $R$-dependent) constant.
However, as it stands \rz sol is not well-defined for loops
intersecting one of the poles $X^i=(0,0,\pm R)$ and also, generically, it is
discontinuous there. (Consider, e.g.,
an infinitesimal loop
winding around the North-pole $X^i=(0,0,R)$ and another one close to
that North-pole having trivial winding number. On the first loop \rz sol
yields (approximately) $\Psi=\exp(i2\pi R)$, on the second one
$\Psi=1$). Multiplying \rz sol by
$\exp (-i R \oint \6 \Phi dx)$, which, on the patch of validity of \rz
sol in $\G_{\CS}$,  is an $R$-dependent constant,
the discontinuity of \rz sol at the
North-pole may be removed. However, continuity at the South-pole
yields a restriction to loops on spheres $\CS$ with integer or half-integer
radius $R$.

The situation is overviewed best \cite{Amati}, if one reformulates \rz
sol with Stokes' theorem as
\be  \Psi[\CX] \propto \exp\left(i\int_D dZ \wedge d\Phi \right) \; ,
\quad \6 D = \CX  \; , \eel symplsol
i.e.\ $D$ is a two-surface the boundary of which is the argument loop
of the wave functional. Now, the two-form $dZ \wedge d\Phi=: \O$ is
well-defined all over a two-sphere $\CS$ of fixed radius $R$; obviously $\O$
is nothing but the symplectic two-form on $\CS$. \
According to \rz symplsol  the
phase factor of $\Psi$ is the area
enclosed by $\CX$ (measured  by means of $\O$). All the ambiguity is
now separated into the ambiguity of choosing  $D$ on a
two-sphere $\CS$ such that its boundary coincides with a given loop $\CX$.
Obviously there are two  inequivalent choices for $D$. $\Psi$ should
be unaffected by this choice. This is the case, iff  the total
area of the sphere is an integer multiple of $2\pi$, which again leads
to $2R \in Z$.

In this formulation the generalization to
arbitrary Poisson $\s$-models is straightforward: For any integral
symplecitc leaf there exists one quantum state of the Poisson
$\sigma$-model. By definition a  symplectic leaf $\CS$
is called integral, iff the integral of the symplectic form $\O$
 over any closed surface $\S \subset \CS$ is an integer multiple of
 $2\pi$:
\be \oint_\S \O = 2 \pi n \; , \quad n \in Z \; . \eel integr

To be more precise, this result only holds, if the first homotopy of
the symplectic leaves $\CS$ is trivial. Otherwise it might happen, that a
loop $\CX$ serving as argument in the wave function does not enclose an
area. In this case we have to take the area between $\CX$ and a
reference loop in the same homotopy class as the phase factor in the
wave function. (The construction given above for trivial
homotopy corresponds to choosing a constant (pointlike) loop as
reference loop).
Effectively this yields that the wave function depends on an additional
quantum number parametrizing the first homotopy group of the
symplectic leaves.

To get further insight into the integrality condition \ry integr ,
observe that the integrality of the symplectic
form on a classical phase space is nothing  but
the Bohr-Sommerfeld quantization
condition. So the integrality condition \rz integr means that the classical
phase space associated with a symplectic leaf is
quantizable \cite{Wood}. We already mentioned that the
symplectic leaves in the
SU(2)-example may be seen as the classical analogues of spin
systems. In that case the integrality condition expresses the fact that spin
is always integer or half integer. According to the analysis above, to
any {\em quantum} spin system there exists {\em one} state of the
$SU(2)$ YM-theory.

Next let us specify the results for the example of Euclidean
$\CR^2$-gravity characterized by the Poisson structure \ry PR2 .
We found that for any value of the Casimir $C_{\CR^2}$ there exists
one symplectic
leaf which is diffeomorphic to $\dR^2 \sim T^\ast \dR$ (beside possibly
additional ones that will be dealt with below). These leaves
are certainly always quantizable phase spaces, \rz integr is
 satisfied trivially. So the proportionality constant in
\rz symplsol becomes an arbitrary function of $C_{\CR^2}$ here.
This leads to
physical wave functions on a line effectively.
However, this is not all of the
story. We know that for $C_{\CR^2} \in [-1/6,1/6]$ there are
further symplectic leaves and we have to find out which of them are
quantizable.

For  the pointlike leaf at $C_{\CR^2}=- 1/6$ the integrality condition
\rz integr is
satisfied trivially. So, we
obtain one additional quantum state of  $\CR^2$-gravity for this
value of the Casimir. (It is arguable, however, whether this state
contributes to the physical Hilbert space: Possibly it should
not be taken into account because the corresponding gauge orbit
is of lower dimension than the generic ones). Next we have to
clarify which of the elliptic leaves are quantizable ($C_{\CR^2}
\in (-1/6,1/6)$).  This is not too difficult: Integrating the
symplecitc form $\O=dZ \wedge d\Phi$ over the respecitve
ellipsoid, obviously one obtains an integer multiple of $2\pi$,
iff the 'height' $Z_{max}-Z_{min}$ of the ellipsoid is an
integer. An explicit calculation shows that this is the case
only for the ellipsoid at $C_{\CR^2}=1/9\sqrt{6}$ ($\ra
Z_{max}=1/2 -1/\sqrt{6}$, $Z_{min}=-1/2 -1/\sqrt{6}$). This
definitely provides an additional quantum state of
$\CR^2$-gravity and the spectrum of the Casimir becomes twofold
degenerate at that specific value.  At $C_{\CR^2}=1/6$ the
situation is not that clear: There are three leaves, one of
which is not simply connected. The latter fact would lead to an
infinite number of states at that value of $C_{\CR^2}$. However,
the space of gauge orbits is not Hausdorff precisely there. Some
continuity requirement thus might lead effectively to an
identification of all these additional states. Evidently, at this point
there is some ambiguity in how to define a quantum theory.

Studying \rz R2acti for Minkowskian metrics $g$, one again obtains a
continuous spectrum for $C_{\CR^2}$. Now, however, this spectrum is
infinitely degenerate within all of the range $(- 1/6,1/6)$. The reason
 is that those leaves that have been elliptic in the Euclidean
formulation of the theory, turn into (two-dimensional)  non-compact
($\Ra$ condition \rz integr statisfied always) and
non-simply connected ones (cf.\ \cite{Klo} for more details).

As a last application of the theory of Poisson $\s$-models we consider the
$G/G$ WZW model. Its action is
\be
  \begin{array}{r@{}l}
   L_{GWZW}&(g,a_+,a_-)= {k \0 8 \pi} \int_M  tr\, \left(
\dot g g^{-1}\dot gg^{-1}
-   \partial gg^{-1}\partial gg^{-1}\right) \, d^2x +
    \\[2pt] &+       {k \0 12 \pi} \int_{B\,\mbox{with} \,
     \partial B=M} \; tr \, \left(dgg^{-1} \wedge
    dgg^{-1} \wedge dgg^{-1}\right) \\[2pt]
                   &+ {k \0 4  \pi}  \int_M tr
                \, [a_+(\dot g - \6 g) g^{-1}
                    -a_- g^{-1}                    (\dot g + \6 g)
                    -a_+ga_- g^{-1} +a_+a_-] \, d^2x \, ,
  \end{array}
 \eel agwzw
where $g$ takes values in some  Lie-group $G$ and
$a_\pm$  in the corresponding Lie-algebra.
The second line, the Wess-Zumino part of the action,
 requires a restriction of $k$ to integers;
under this condition it  becomes single-valued modulo $2\pi$.
The first two lines in \rz agwzw are a conformal
generalization of the action for a free scalar field: Indeed
for $g \in U(1)$ $\lra g=\exp(i\phi)$ this action  becomes
proportional to $\int [\dot \phi^2 - (\6 \phi)^2]d^2x$.
The third and last
line in \rz agwzw ensures that the global symmetry under adjoint
transformations, $g \ra h^{-1}gh$, of the previous lines
becomes a local gauge symmetry  of  the action (upon
$a_\pm \ra h^{-1} a_\pm h + h^{-1} \dot h \pm  h^{-1} \6 h$).

For semisimple $G$ \rz agwzw may be brought into Poisson
$\s$-form. This shall be  displayed for $G=SU(2)$. In that case
\be g=\left(
  \begin{array}{rr}
  X^1+iX^2&X^3+iX^4\\-X^3+iX^4 & X^1 - i X^2
  \end{array}
  \right) \; ,
\eel su2
where the $X^i \in \dR$ are subject to the condition
\be \sum_{i=1}^4 X^iX^i =1 \; . \eel cond
This  displays best that  topologically $SU(2)$ is a
three-sphere $S^3$. Also it is easy to determine the symmetry
orbits  of \ry agwzw , i.e.\ the conjugacy classes $g \sim
h^{-1} g h$: They are described by constant values of $X^1= tr \, g /2$ and
according to \rz cond they are two-spheres for $X^1 \in (-1,1)$  and
points for the poles of the  $S^3$ $X^1=\pm 1$.

Rewriting \rz agwzw in first order form and integrating out $a_+$, which
then enters quadratically, one finds \cite{Anton}
\be L'_{GWZW}=L_{top}(g,A)+ L_\d(g)  \, . \eel neu
Here $L_{top}$ is the action \rz acti2 with Poisson bracket
\be
\{ X^i, X^j \} ={2 \pi \0 k} \, (X^1 + i X^2) \, \sum_{l=1}^4
\e^{1ijl} X^l \, , \eel psu2
where $\e^{ijlm}$ is the completely antisymmetric four-tensor.
$L_\d$, on the other hand, is
\be L_\d = k\int_B \delta(X^1)\delta(X^2) \, dX^1 \wedge dX^2
\wedge d\phi \; ,
\eel ldelta
where we have introduced an angular variable $\phi$
via
\be  \exp(i\phi) := {X^3 + i X^4 \0 \sqrt{(X^3)^2
+ (X^4)^2}} \; .  \eel coord
$L_\d$ is  a relict of the Wess-Zumino term in \ry agwzw .
The corresponding  integrands
\be \delta(X^1)\delta(X^2) \, dX^1 \wedge dX^2
\wedge d\phi   \eel coho
and $dgg^{-1} \wedge     dgg^{-1} \wedge dgg^{-1} / 12 \pi$,
respecitvely, are from the same (non-trivial) cohomology class
on $SU(2)$ and  \rz coho is a representative of this class with minimal
support. The support of  \rz coho is merely  an  $S^1$ ($X^1=X^2=0
\Ra (X^3)^2+(X^4)^2 = 1$), which obviously is contained entirely
within the conjugacy class $X^1=0$. As a consequence we
will be  allowed to disregard $L_\d$
in the analysis of \rz neu for all symmetry orbits $X^1 \neq 0$.

The Poisson tensor \rz psu2 is not real and
consequently also the $A_i$ are
complex valued now. Still the equations of motion resulting from the
action \rz neu as well as the Hamiltonian structure
are completely equivalent to the
original formulation \ry agwzw , at least if
appropriate conditions for $A$  at $(X^1,X^2) \to (0,0)$ are
taken into account, cf.\ \cite{Anton}. As  the
  complex conjugate fields $\bar A_i$ do not enter \ry neu ,
the $\bar A_i$ may be eliminated  in a  Hamiltonian formulation
right at the beginning and, for $X^1 \neq 0$, one is left again with
 \rz const and \ry poibr .

Let us remark here also that due to \rz cond
the $dX^i$ are not linearly independent. Consequently
only three of the $A_i$ are  independent. Alternatively we  could
have introduced  true local coordinates $X^i$ on $SU(2)$, $i=1,2,3$.
However,
 the embedding of $SU(2)$ into $\dR^4$ has advantages with
respect to  a global
description. Note that  the restriction of the Poisson
tensor \rz psu2 on $\dR^4$
to $SU(2)$ is consistent, because the left-hand side of \rz cond is a
Casimir function.

As expected $X^1$ is a Casimir function of \ry psu2 . However, the
direct identification of  conjugacy classes  of $SU(2)$ with
symplectic leaves fails for the two-sphere $X^1=0$, as obviously
the Poisson tensor $\CP^{ij}$ vainishes at its  equator
$(X^1,X^2) = (0,0)$.
This artifact turns out to be cured, however, when
$L_\d$ and the boundary conditions for the $A$-fields are taken into account
\cite{Anton}, so that also $X^1=0$ corresponds to one symmetry
orbit of \rz neu only.

Let us determine which of the symplectic leaves are integral now. Certainly
the pointlike ones at $X^i=(\pm 1, 0,0,0)$ are.
As the angular variable $\phi$ and
$(ik/2\pi)\, \ln (X^1 + i X^2)$ are canonically conjugates with
respect to \ry psu2 , on the remaining
two-spheres $X^1=const \neq 0$ the symplecitc two-form  may be written as
\be \O_{GWZW} = {ik \0 2 \pi} d \ln (X^1 + i X^2)
\wedge d \phi \, . \eel sym
Introducing another spherical coordinate $\theta$ on $S^3$ by
$X^1:= \cos \theta$, $\theta \in [0, \pi]$,
the respective $S^2$ is parametrized  by $\phi \in [0, 2\pi]$ and $X^2 \in
[-\sin \theta,\sin \theta]$, with $\theta$ fixed,  and the integral over the
symplectic form becomes
\be \oint_{S^2}  \O_{GWZW} =  \left\{ \begin{array}{r@{\quad , \quad}l}
2 k \theta & \theta \in \, [0,\pi/2) \; \lra \; X^1 \in (0,1]
\\ 2 k (\theta - \pi) &  \theta \in
\, (\pi/2,\pi]  \; \lra X^1 \in [-1,0)  \quad .
\end{array} \right. \eel volume
Here we made use of the fact that for $X^2=\pm \sin \theta$ one  has
$\ln (X^1 + i X^2) = 2i \theta + n 2 \pi i$, where the
integer $n$ is determined most
easily by the requirement that $\int   \O_{GWZW} =0$ for
the pointlike leaves at
$X^1 \equiv \cos \theta = \pm 1$. So, those leaves characterized by a
value of $X^1=\cos \theta$ yielding an integer multiple of $2\pi$ for
the righthand side of \rz volume  carry a quantum state of the
$SU(2)$-GWZW-model.

At the equator of the two-sphere $X^1=0 \lra \theta = \pi/2$
clearly the symplecitc form  \rz sym becomes ill-defined and
so becomes its integral
\ry volume . Here one has to take into account the changes
induced by $L_\d$ and the above-mentioned conditions on the $A$-fields.
However, the correct result  may be guessed already by a
simple limiting procedure:
{}From \rz volume we obtain
\be
\lim_{X^1 \to 0}\; \, \int_{S^2}\Omega_{GWZW} =  k\pi
 \qquad \mbox{mod} 2\pi \, ,
\eel aa
which, for integer $k$, is unique up to $2 \pi$. As may be
verified by a more
careful analysis \cite{Anton}, indeed
the  adjoint orbit $X^1=tr \, g /2=0$  carries a quantum state,
iff \rz aa is
an integer multiple of $2\pi$, i.e.\ iff  $k$ is even.

Summing up, we conclude that the integral orbits (i.e.\ the
orbits or conjugacy classes
allowing for  nontrivial quantum states of the $SU(2)$-GWZW model)
are given by $\theta = n\pi/ k  \lra X^1= \cos (n\pi/ k)$, $n=0,1,... k$ .
Here we counted also the two states localized at the pointlike orbits
at $g=\pm \done$. If one
disregards these states, one should renormalize $k$ to
$k+2$ simultanously  so as to reproduce standard path integral results.

It is instructive to compare the final picture with that of the YM theory.
There  we had $N=su(2)=\dR^3$, now we have $N=SU(2)=S^3$.
Both target spaces $N$ are foliated
into two-spheres, one respectively two of which degenerate to a point.
In the YM-case an infinite number of these two-spheres is integral,
in the G/G case it is a finite number only.
We have seen, furthermore, that any of the YM-quantum states corresponds
to a quantum spin system and thus to
a unitary representation of the $su(2)$-algebra.
One can show that similarly any of the
G/G-quantum states is  associated
to  a representation of the quantum group
$su_q(2)$  with $q=\exp[i2\pi/(k+2)]$.
 These results generalize  to arbitrary compact
gauge groups  in an obvious manner.

A final remark:
The YM-theory taught us a lecture
of what kind of terms may be added to $L_{top}$ without spoiling its
symmetries \ry symme .
So does the $G/G$-model: Given a two-form or a
closed three-form on $N$ which is invariant under all symmetries
generated by
vector-fields of the form $\CP^{ij} d/dX^j$ (as is the case for
\ry coho ), one may extend the Lagrangian  by adding the
pull-back  of this form to $M$. As this does not involve any metric
or volume-form on $M$, the resulting action is still
topological.

Already by means of these simple examples the richness of the
Poisson $\s$-model on the quantum level becomes obvious. Still with
respect to the calculation of the partition function for the general
model only partial results have been achieved yet \cite{proce}. It will
be interesting to investigate further what kind of topological features
of the target space of the theory are encoded in the
partition function and
correlators of the Poisson $\s$-model. Independently  of this,
it  should have
become clear by the exposition above that  Poisson $\s$-models
allow for a
comprehensive treatment of many seemingly different 2D field theories.

\section*{Acknowledgement}
We are grateful to A.Alekseev for collaboration on some of the topics
presented here. Moreover we want to thank the organizers of the
school for giving us the opportunity for this presentation,
T.Asselmeyer for
discussions,  and H.Pelzer for critically reading the manuscript.

\end{document}